\documentclass[12pt,prd,aps,onecolumn,nofootinbib]{revtex4-1}%
\usepackage[colorlinks=true,linkcolor=blue,urlcolor=blue,filecolor=black,citecolor=red,pdfstartview=FitV,pdftitle={},pdfsubject={},
pdfkeywords={},pdfpagemode=None,bookmarksopen=true]{hyperref}
\usepackage{graphicx}
\usepackage{epstopdf}%
\usepackage{amsmath}
\usepackage{amsfonts}
\usepackage{amssymb}
\usepackage{xcolor}
\usepackage{color}%
\usepackage{dcolumn}
\usepackage{slashed}
\usepackage[normalem]{ulem}
\usepackage{amsthm}
\usepackage{mathrsfs}
\usepackage{subfigure}
\usepackage{wrapfig}
\usepackage{indentfirst}

\providecommand{\U}[1]{\protect\rule{.1in}{.1in}}

\begin{document}

\title{Spontaneous scalarization of regular Hayward black holes in Einstein-nonlinear electromagnetic-scalar gravity}

\author{Lan-Lan Cai$^{1}$, Meng-Yun Lai$^{1}$\footnote{mengyunlai@jxnu.edu.cn (corresponding author)}, De-Cheng Zou$^{1}$\footnote{dczou@jxnu.edu.cn}, Lina Zhang$^2$\footnote{linazhang@hnit.edu.cn}\\ and Hyat Huang$^{1}$\footnote{hyat@mail.bnu.edu.cn}}


\affiliation{$^{1}$College of Physics and Communication Electronics, Jiangxi Normal University, Nanchang 330022, China\\
$^{2}$College of Science, Hunan Institute of Technology, Hengyang 421002, China}

\begin{abstract}
\indent
Regular Hayward black holes provide a useful setting for investigating scalarization in theories with nonminimally coupled matter sectors. Within the framework of Einstein-nonlinear electromagnetic-scalar gravity, we identify the tachyonic threshold that signals the bifurcation from the bald Hayward background and then obtain scalarized charged black holes for both quadratic $(1-\alpha\phi^2)$ and exponential $(e^{-\alpha \phi^2})$ couplings. These configurations form a discrete set of branches classified by the number of nodes in the scalar field. The branch with $n=0$ is the fundamental branch, whereas solutions with $n\geq 1$ are excited branches. By studying radial perturbations, we find that the fundamental branch is stable for both coupling choices, which makes it the most relevant branch for future phenomenological and observational studies.
\end{abstract}

\maketitle

\section{Introduction}

Classical general relativity predicts that gravitational collapse can culminate in spacetime singularities \cite{Hawking:1973}, a result commonly interpreted as indicating the limits of validity of the classical theory. This has motivated the search for black hole geometries in which curvature invariants remain finite everywhere. Among these constructions, regular black holes are particularly important because they preserve the basic external properties of black holes while modifying the short-distance structure of spacetime.

A well-known prototype is the Bardeen black hole \cite{Bardeen:1968}, which was later associated with nonlinear electrodynamics as the source of a magnetically charged regular configuration. Hayward subsequently proposed another regular solution \cite{Hayward:2005gi}, characterized by static spherical symmetry and the absence of a central singularity. The corresponding metric is~\cite{Fan:2016rih,Fan:2016hvf,Huang:2025uhv}
\begin{eqnarray}
   && ds^2 = -f(r)dt^2 + f(r)^{-1}dr^2+r^2(d\theta^2+\sin^2\theta d\varphi^2),\nonumber\\
     && f(r)=1-\frac{2Mr^2}{r^3+q^3},\label{Hayward solution}
\end{eqnarray}
with the magnetic field configuration
\begin{eqnarray}
F_{\theta\varphi} = q\sin\theta, \quad with \quad {A}_{\varphi } = -q\cos\theta .
\end{eqnarray}
Here $M$ and $q$ are the mass and magnetic charge parameters. More generally, nonlinear electrodynamics provides a natural mechanism for supporting several classes of regular black holes, as emphasized in the work of Ayon-Beato et al. \cite{Ayon-Beato:1998,Ayon-Beato:2000}. Consequently, these solutions have attracted sustained attention in studies of black hole interiors, effective matter sources, and possible modifications of strong-gravity physics \cite{Huang:2020,Zou:2021,Berej:2006cc,Kumara:2021hlt}.

Spontaneous scalarization provides a mechanism through which compact objects can acquire scalar hair only in sufficiently strong gravitational environments, while remaining compatible with weak-field constraints~\cite{Damour:1993,Damour:1996}. The basic ingredient is a nonminimal coupling that renders the effective scalar mass squared negative in an appropriate background, thereby destabilizing an initially bald configuration. Once the instability is triggered, nonlinear effects can eventually saturate the growth and support new scalarized equilibrium states. This mechanism has been established in several classes of gravitating systems. In Einstein-scalar-Gauss-Bonnet theories, the onset of scalarization is associated with the destabilization of the Schwarzschild background through the Gauss-Bonnet coupling \cite{Doneva:2017bvd,Antoniou:2017acq,Silva:2017uqg}. Closely related phenomena also arise in Einstein--Maxwell--Scalar models for Reissner--Nordstr\"om black holes, including rotating solutions and spin-induced scalarization \cite{Herdeiro:2018wub,Fernandes:2019rez,Myung:2018vug,Zou:2019bpt,Astefanesei:2019pfq,Cunha:2019,Collodel:2020,Dima:2020,DonevaPRD:2020,DonevaEPJC:2020,Herdeiro:2021,Berti:2021}. Analogous scalarization mechanisms have furthermore been reported for boson stars, charged stars, and horizonless reflecting compact objects in scalar-tensor settings \cite{Whinnett:2000,Brihaye:2019,Minamitsuji:2021,PengJHEP}.

Building on our earlier study of scalarization in regular magnetically charged black holes~\cite{Zhang:2024bfu}, we now turn to the Hayward solution~\cite{Fan:2016rih,Fan:2016hvf,Huang:2025uhv}, which provides another important regular black hole background. Our goal is to determine whether Hayward black holes can also undergo spontaneous scalarization and to clarify the properties of the resulting scalarized configurations for two representative coupling models.

This paper is organized as follows. In Section~\ref{2s}, we introduce the underlying Einstein-nonlinear electrodynamics-scalar theory and identify the instability threshold of the bald Hayward background. Section~\ref{3s} presents the numerical construction of the fundamental scalarized branch. In Section~\ref{4s}, we examine its radial stability. We conclude in Section~\ref{5s} with a summary and discussion.

\section{Instability for Hayward black holes}
\label{2s}

To analyze scalarization in the Hayward background, we consider an Einstein-nonlinear electrodynamics system supplemented by a nonminimally coupled scalar field, with action
\begin{eqnarray}
I=\frac{1}{16\pi}\int d^4x\sqrt{-g}\left[R-2(\nabla\phi)^2-4\xi(\phi)L(\mathcal{F})\right],
\end{eqnarray}
where $R$ denotes the Ricci scalar, $\phi$ is a real scalar field, and the interaction between the scalar and electromagnetic sectors is encoded in the coupling function $\xi(\phi)$, which multiplies the nonlinear electrodynamics (NLED) Lagrangian $L(\mathcal{F})$. Physically, it represents an effective electromagnetic matter sector whose stress-energy tensor regularizes the central region. For the Hayward regular black hole, the corresponding NLED Lagrangian is given by ~\cite{Benavides-Gallego:2018odl,Nam:2018uvc,Li:2025glq}
\begin{eqnarray}\label{LF}
L(\mathcal{F})=\frac{3M}{q^3}\frac{(2q^2\mathcal{F})^{3/2}}{\left[1+(2q^2\mathcal{F})^{3/4}\right]^2},
\end{eqnarray}
where $\mathcal{F}=F^{\mu\nu}F_{\mu\nu}/4$, $M$ and $q$ denote the mass and magnetic charge parameters, respectively.

Varying the action with respect to $g_{\mu\nu}$, $\phi$, and $A_\mu$, we obtain the Einstein, scalar, and generalized Maxwell equations,
\begin{eqnarray}
&&R_{\mu\nu}-\frac{1}{2}g_{\mu\nu}R = 4\xi(\phi)\left[2\frac{\partial L(\mathcal{F})}{\partial \mathcal{F}}F_{\mu\rho}F_{\nu}^{~\rho}-\frac{1}{2}g_{\mu\nu}L(\mathcal{F})\right]
+2\partial_\mu\phi\partial_\nu\phi-(\nabla\phi)^2g_{\mu\nu}, \label{EinsteinEq}\\
&&\nabla^2\phi -\frac{\partial \xi(\phi)}{\partial \phi}L(\mathcal{F})=0, \label{scalarEq}\\
&&\nabla_\mu\left[4\xi(\phi)\frac{\partial L(\mathcal{F})}{\partial \mathcal{F}}F^{\mu\nu}\right]=0. \label{MaxwellEq}
\end{eqnarray}

 To explore the onset of scalarization, we consider the following two scalar coupling functions:
\begin{eqnarray}\label{couplingf}
\xi(\phi)=
\left\{
\begin{array}{ll}
1-\alpha\phi^2, & \mbox{quadratic coupling},\\
e^{-\alpha\phi^2}, & \mbox{exponential coupling}.
\end{array}
\right.
\end{eqnarray}
The bald Hayward solution is recovered in the scalar-free sector $\phi=0$, where the remaining equations reduce to those of the underlying nonlinear electrodynamics system.
Although these two couplings differ at the nonlinear level, they lead to the same linearized scalar perturbation equation around the bald background, and their differences only appear through higher-order terms. 

Neglecting metric perturbations, the scalar perturbation $\delta\phi$ on the Hayward background satisfies
\begin{eqnarray}
\bar{\nabla}^{2}\delta\phi-\mu_{\rm eff}^{2}\delta\phi=0,\qquad
\mu_{\rm eff}^{2}=-2\alpha L(\mathcal{F}),
\end{eqnarray}
Here $\bar{\nabla}_\mu$ denotes the covariant derivative associated with the fixed bald Hayward background, and barred quantities are evaluated on this background.
Then, the coupling generates an effective mass squared determined by the nonlinear electromagnetic background. When $\mu_{\rm eff}^{2}$ becomes sufficiently negative outside the horizon, tachyonic growth may arise, indicating an instability of the bald Hayward black hole.

To construct these solutions, the general static and spherically symmetric line element can be written as
\begin{eqnarray}\label{metric}
ds^2=-N(r)e^{-2\delta(r)}dt^2+\frac{dr^2}{N(r)}
+r^2(d\theta^2+\sin^2\theta d\varphi^2),
\end{eqnarray}
with
\begin{eqnarray}
N(r)=1-\frac{2m(r)}{r}, \qquad \phi=\phi(r).
\end{eqnarray}

Considering  the scalar-free configuration $\phi=0$, one has $\xi(0)=1$ and $\xi_{,\phi}(0)=0$ for both choices in Eq.\eqref{couplingf}. From Eqs.~\eqref{EinsteinEq}, \eqref{scalarEq} and \eqref{MaxwellEq}, the generalized Maxwell equation is satisfied with the magnetic monopole $A_\varphi=-q\cos\theta$. The remaining Einstein equation gives
\begin{eqnarray}
m'(r)=\frac{3Mq^3r^2}{(r^3+q^3)^2},
\end{eqnarray}
which integrates to $m(r)=Mr^3/(r^3+q^3)$. Hence, $N(r)=1-2m(r)/r=f(r)$, as shown in Eq.\eqref{Hayward solution}.

As a representative nonextremal example, we take $M=0.5$ and $q=0.25$, for which the outer horizon is located at $r_+=0.935$ from $f(r_+)=0$. To analyze the linear scalar response of the Hayward background, we decompose the perturbation as
\begin{eqnarray}
\delta \phi (t, r, \theta, \varphi) = \frac{u (r)}{r} e^{-i\omega t}Y_{lm} (\theta, \varphi).
\end{eqnarray}
Introducing the tortoise coordinate $r_*$ through
\begin{eqnarray}
r_*=\int \frac{dr}{f(r)},
\end{eqnarray}
the scalar perturbation equation can be cast into the Schr\"odinger-like form
\begin{eqnarray}
\label{perturbedEq}
\frac{d^2u}{dr_*^2} + \left(\omega^2 - V(r)\right)u(r) = 0,
\end{eqnarray}
with the effective potential
\begin{eqnarray}
V (r) = f (r)
\left[
\frac{l(l + 1)}{r^2}
+\frac{6Mr^3}{(q^3 + r^3)^2}
-\frac{4M}{q^3 + r^3}
-\frac{6q^3M\alpha}{(q^3 + r^3)^2}
\right].
\end{eqnarray}

\begin{figure}[t!]
   \centering
\includegraphics[width=0.5\textwidth]{ 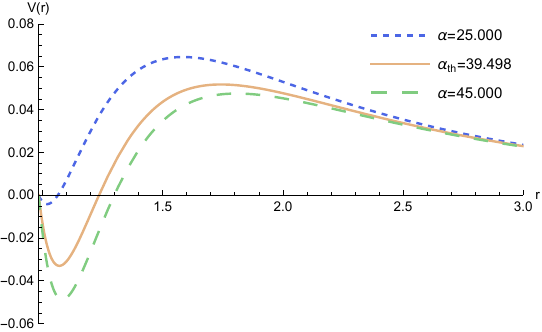}
\caption{Plots of the potential $V\left( {r,\alpha ,q}\right)$ for three different values $\alpha=\left\{{{25},{\alpha}_{\text{th}} = {39.498},{45}}\right\}$ are shown from top to bottom near the $V$ axis,
with ${\alpha }_{\mathrm{{th}}}\left( q\right)={39.4976}\left( {0.250}\right), {21.1872}\left({0.300}\right), 12.0436\left( {0.350}\right)$.}\label{potV1fig}
\end{figure}

To determine whether scalarized solutions can bifurcate from the bald Hayward branch, we first examine the linear instability of the background. Here, we focus on the spherically symmetric mode $l=0$, for which the corresponding potential is shown in Fig.~\ref{potV1fig}. Although a positive-definite potential outside the horizon is sufficient to ensure mode stability, the presence of a negative well near the horizon only suggests the possibility of tachyonic growth and does not, by itself, prove instability. The decisive criterion is whether the perturbation spectrum contains modes that grow in time. For this reason, we solve Eq.~\eqref{perturbedEq} numerically after setting $\omega=-i\Omega$, with purely ingoing behavior at the horizon and purely outgoing behavior at spatial infinity.

To determine the instability threshold, it is essential to identify $\alpha_{\rm th}(q)$, which marks the onset of scalarization. This can be achieved by searching for static bound-state scalar configurations (scalar clouds) of Eq.~\eqref{perturbedEq} on the Hayward background, obtained by setting $\Omega=0$ and writing $u(r)=r\psi(r)$. For the spherically symmetric mode $l=0$ with $M=1/2$, the requirement of asymptotic normalizability selects a discrete set of critical values $\alpha_n(q)$, where $n=0,1,2,\ldots$ counts the number of nodes of the scalar-cloud profile $\psi(r)$.

\begin{figure*}[t!]
\centering
\subfigure[$\Omega\sim \alpha$]{\includegraphics[width=0.4\textwidth]{ 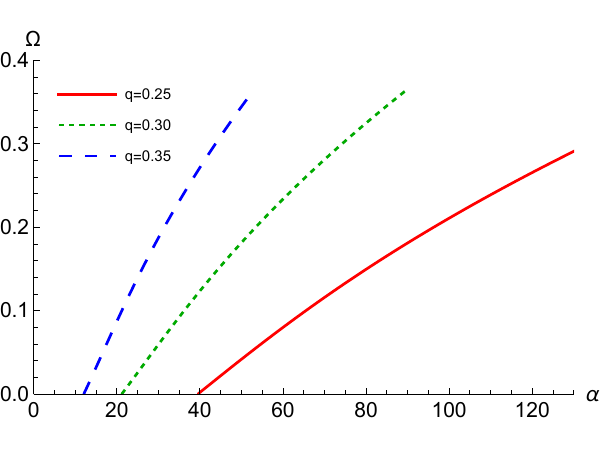}\label{eq2a}}
\qquad
\subfigure[$\psi\sim z$]{\includegraphics[width=0.4\textwidth]{ 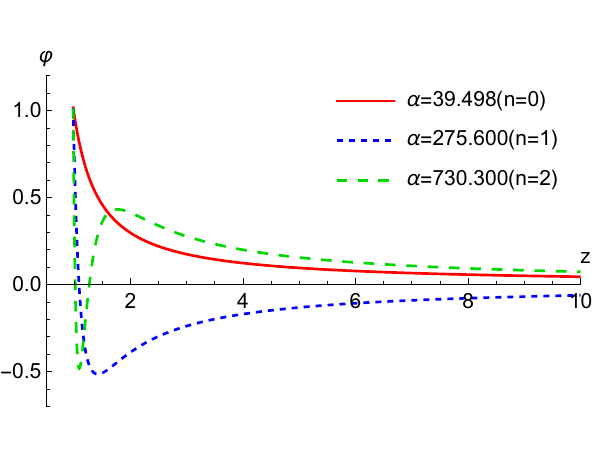}}
\caption{(a) Three curves of $\Omega$ in ${e}^{\Omega t}$ as functions of $\alpha$ are used to determine the instability thresholds ${\alpha }_{\mathrm{{th}}}\left( q\right)$ around a Hayward black hole. We find ${\alpha }_{\mathrm{{th}}}\left( q\right)= {39.4976}\left( {0.25}\right)$, ${21.1872}\left( {0.30}\right)$, and $12.0436(0.35)$ when the three curves cross the $\alpha$ axis. 
(b) Radial profiles as functions of $z = r/{2M}$ for $M = {1/2}$ and $q = {0.25}$, showing the first three static scalar perturbation solutions. The number $n$ of nodes labels the $n=0,1,2$ scalarized charged black hole (SCBH) solutions.}\label{fig2}
\end{figure*}

Figure~\ref{fig2} displays the first few scalar clouds as functions of $z=r/(2M)$. The node-free mode $n=0$ gives rise to the fundamental scalarized branch, whereas the modes with $n\geq 1$ correspond to excited branches. More generally, each scalar cloud seeds a corresponding branch of scalarized charged black holes, so that $\{\alpha_0,\alpha_1,\alpha_2,\ldots\}$ can be interpreted as the bifurcation points of these branches. In particular, for a given magnetic charge $q$, the instability threshold is determined by
\begin{eqnarray}
\alpha_{\rm th}(q)=\alpha_{n=0}(q),
\end{eqnarray}
which identifies the onset of the dominant fundamental branch.

\begin{figure}[htbp]
\centering
\includegraphics[width=0.5\textwidth]{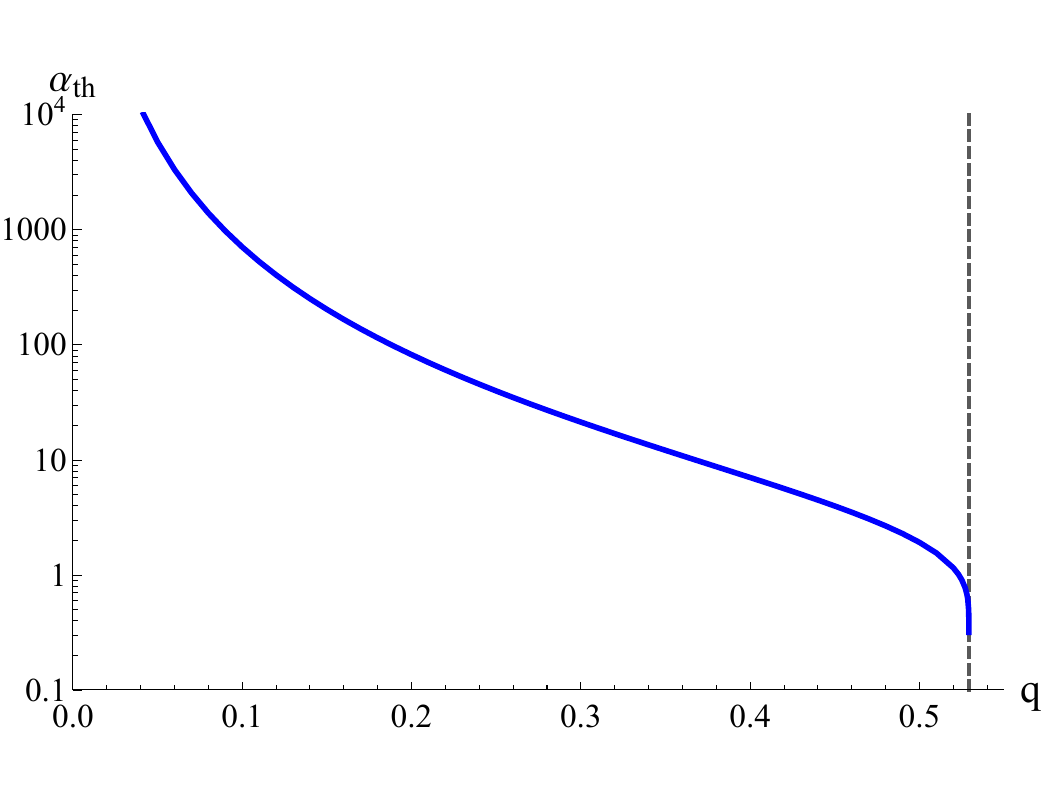}
\caption{
Threshold coupling $\alpha_{\rm th}(q)$ for $M=1/2$. 
The curve decreases monotonically with $q$, diverging rapidly as $q\to 0$ and approaching the weak-coupling regime near $q_c\simeq0.5291$.
}
\label{fig:alpha_th_q}
\end{figure}

Unlike linear Maxwell electrodynamics, the present NLED model depends explicitly on the Hayward parameter $q$ through $L(F)$. Consequently, the scalar-cloud eigenvalue problem does not admit a simple analytic scaling relation for the critical coupling $\alpha_{n=0}$. We therefore determine $\alpha_{n=0}$ numerically for each value of $q$. For $M=1/2$, the three representative values are
\begin{eqnarray}
\alpha_{n=0}=39.4976,\;21.1872,\;12.0436
\quad \text{for} \quad
q=0.25,\;0.30,\;0.35,
\end{eqnarray}
respectively. The corresponding threshold curve is shown in Fig.~\ref{fig:alpha_th_q}. 
The critical coupling decreases monotonically as the magnetic charge $q$ increases. 
In the small-$q$ regime, $\alpha_{\rm th}$ grows rapidly, indicating that the scalarized branch is pushed to very large coupling as $q\to 0$. 
On the other hand, as $q$ approaches the maximal Hayward value
$q_c\simeq 0.5291$, the threshold coupling is driven to the small-coupling regime.

\section{SCALARIZED CHARGED BLACK HOLES}
\label{3s}

Scalarized charged black holes (SCBHs) bifurcate from the corresponding scalar clouds in the unstable regime of the Hayward background, i.e., for $\alpha(q)\geq \alpha_{\rm th}(q)$. 

For the ansatz \eqref{metric}, the generalized Maxwell equation is solved by the same magnetic potential as in the bald Hayward case,
\begin{eqnarray}
A_\varphi=-q\cos\theta,
\end{eqnarray}
which gives
\begin{eqnarray}
F_{\theta\varphi}=q\sin\theta, \qquad \mathcal{F}=\frac{q^2}{2r^4}.
\end{eqnarray}
Therefore, the magnetic potential field is not affected by scalarization and does not introduce an additional function in the numerical construction. In fact, for a static and spherically symmetric scalar field, $\xi(\phi)$ depends only on $r$, and the generalized Maxwell equation \eqref{MaxwellEq}
\begin{eqnarray}
\nabla_\mu\left[\xi(\phi)\frac{\partial L(\mathcal{F})}{\partial \mathcal{F}} F^{\mu\nu}\right]=0
\end{eqnarray}
is automatically satisfied by $F_{\theta\varphi}=q\sin\theta$. Therefore, the scalar field backreaction only modifies the metric functions and scalar profile, but it does not introduce an additional magnetic function.

In what follows, we focus on the fundamental branch $n=0$, which emerges once $\alpha(q)$ exceeds the threshold value $\alpha_{\rm th}(q)$. For the two coupling models,
\begin{eqnarray}
\xi(\phi)=1-\alpha\phi^2, \qquad {\rm and} \qquad \xi(\phi)=e^{-\alpha\phi^2},
\end{eqnarray}
we numerically construct the corresponding scalarized charged black holes for $M=1/2$ and $q=0.25$. Higher-node branches can be obtained in the same manner by starting from the appropriate scalar clouds.

Substituting the static spherically symmetric ansatz into Eqs.~\eqref{EinsteinEq} and \eqref{scalarEq}, we obtain a coupled system of three ordinary differential equations for the functions $\delta(r)$, $m(r)$, and $\phi(r)$:
\begin{eqnarray}
\delta '(r)+r\phi'^2(r)=0,
\end{eqnarray}
\begin{eqnarray}
\frac{6q^3Mr^2\xi(\phi)}{(q^3+r^3)^2}+r(r-2m)\phi'^2(r)-2m'(r)=0,
\end{eqnarray}
\begin{eqnarray}
r(r-2m)\phi''(r)-\left\{m[2-2r\delta'(r)]
+r[2m'(r)+r\delta'(r)-2]\right\}\phi'(r)
-\frac{3q^3Mr^2\xi'(\phi)}{(q^3+r^3)^3}=0,
\end{eqnarray}
where a prime denotes differentiation with respect to $r$.

To construct black hole solutions with an event horizon at $r=r_+$, we impose a regular near-horizon expansion of the form
\begin{eqnarray}\label{horizon}
m(r)=\frac{r_+}{2}+m_1(r-r_+)+\cdots,\nonumber\\
\delta(r)=\delta_0+\delta_1(r-r_+)+\cdots,\nonumber\\
\phi(r)=\phi_0+\phi_1(r-r_+)+\cdots,
\end{eqnarray}
from which the first expansion coefficients are found to be
\begin{eqnarray}\label{horicoef}
m_1=\frac{3q^3Mr_+^2\xi(\phi_0)}{(q^3+r_+^3)^2}, \qquad
\delta_1=-r_+\phi_1^2,\qquad
\phi_1=\frac{3q^3Mr_+\xi'(\phi_0)}{(q^3+r_+^3)^3-36q^4M^2r_+^4\xi(\phi_0)^2}.
\end{eqnarray}
The local horizon expansion is characterized by two free parameters, $\phi_0$ and $\delta_0$.
The ansatz metric functions expand at large $r$ with
\begin{eqnarray}\label{inf}
m(r)=M-\frac{Q_s^2}{2r}+\cdots,\qquad
\phi(r)=\frac{Q_s}{r}+\cdots,\qquad
\delta(r)=\frac{Q_s^2}{2r^2}+\cdots,
\end{eqnarray}
where $M$ is the Arnowitt-Deser-Misner mass and $Q_s$ is the scalar charge. Notice that the scalarized solutions constructed here are exterior black-hole solutions. Their regularity is imposed at the event horizon through Eq.~\eqref{horizon} and checked for \(r\ge r_+\). Whether these solutions admit a globally regular extension to the center requires a separate analysis of the interior geometry, and remains an interesting issue for further investigation~\cite{LiLi:2025qgo}.

\begin{figure*}[t!]
   \centering
\subfigure[$\xi\left( \phi \right)  = 1 - \alpha {\phi }^{2}$]{\includegraphics[width=0.4\textwidth]{ 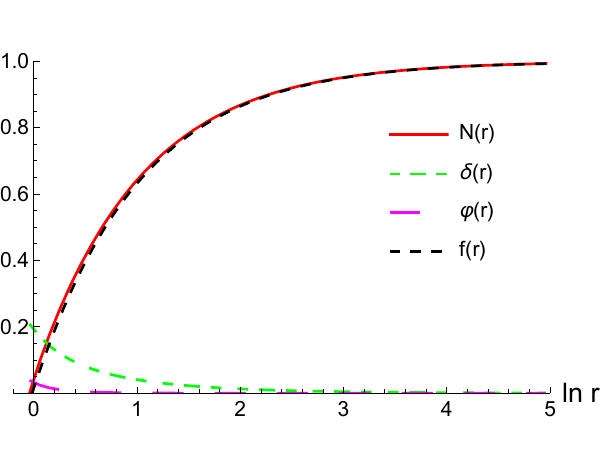}}
\qquad
\subfigure[$\xi\left( \phi \right)  = {e}^{-\alpha {\phi }^{2}}$]{\includegraphics[width=0.4\textwidth]{ 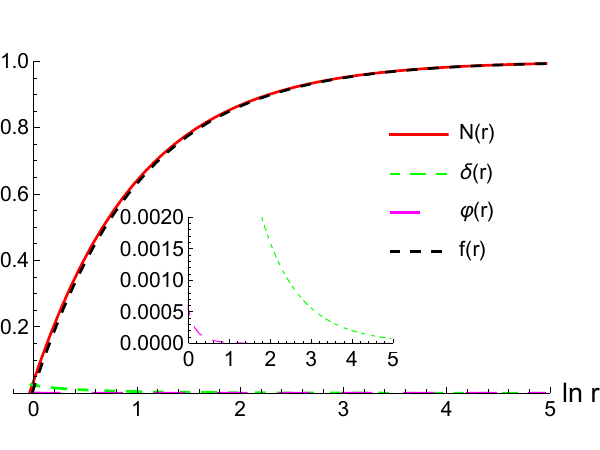}}
\caption{Plots of an SCBH solution with $q= {0.25}$ and $M = {1/2}$ for $\alpha  = {40.8134}$ (quadratic coupling) and $\alpha  = {40.1819}$ (exponential coupling) in the $n = 0$ branch of $\alpha  \geq  {39.4976}$. 
}\label{fig3}
\end{figure*}

For the quadratic coupling, a representative solution on the fundamental branch is shown in Fig.~\ref{fig3}(a) for $\alpha=40.8134$ and $q=0.25$. Compared with the bald Hayward background, the scalarized solution has a slightly shifted horizon location, from $\ln r=-0.0163$ in the Hayward case to $\ln r=-0.0403$ in the scalarized case. Away from the near-horizon region, however, the metric function $N(r)$ approaches the Hayward profile $f(r)$. At the same time, the redshift function $\delta(r)$ decreases with increasing $\ln r$, whereas it vanishes identically for the Hayward solution. The scalar field $\phi(r)$ also decays outward, indicating that the scalar hair is localized mainly in the near-horizon region. An analogous fundamental-branch solution for the exponential coupling is shown in Fig.~\ref{fig3}(b).

\section{STABILITY OF SCALARIZED BLACK HOLES}
\label{4s}

We now investigate the radial stability of the fundamental ($n=0$) branch of scalarized charged black holes. For this purpose, we consider three representative magnetic charges, $q=0.25$, $0.30$, and $0.35$, whose corresponding bifurcation points are $\alpha_{n=0}=\{39.4976,\,21.1872,\,12.0436\}$. We introduce spherically symmetric perturbations around the scalarized background according to
\begin{eqnarray}
ds^2_{\rm RP} &=& -N (r) e^{-2\delta(r)}[1 + \epsilon H_0 (t, r)] dt^2
+\frac{dr^2}{N (r)[1 + \epsilon H_1 (t, r)]}+r^2(d\theta^2+\sin^2\theta d\varphi^2),
\nonumber\\
\phi (t, r) &=& \bar{\phi}(r)+ \epsilon \frac{\delta \phi (t, r)}{r},
\end{eqnarray}
where $N(r)$, $\delta(r)$, and $\bar{\phi}(r)$ describe the background solution, whereas $H_0(t,r)$, $H_1(t,r)$, and $\delta\phi(t,r)$ are linear perturbations. The magnetic gauge field is left unperturbed, and $\epsilon\ll1$ is the perturbative bookkeeping parameter. Since we focus only on radial perturbations, we restrict to the $l=0$ sector. In this case, the physical degree of freedom is fully captured by the scalar perturbation, for which we use the separation
\begin{eqnarray}
\delta \phi (t, r) = \phi_1 (r) e^{\Omega t}.
\end{eqnarray}
The radial perturbation equation can then be reduced to a Schr\"odinger-like form,
\begin{eqnarray}
\frac{d^2\phi_1(r)}{dr_*^2}-\left(\Omega^2+V_{\rm SCBH}(r)\right)\phi_1(r)=0,\label{radialeq}
\end{eqnarray}
where the tortoise coordinate is defined by
\begin{eqnarray}
\frac{dr_*}{dr}=\frac{e^{\delta(r)}}{N(r)},
\end{eqnarray}
and the corresponding effective potential is given by
\begin{align}
V_{\text{SCBH}}(r) &= \frac{{e}^{-{2\delta }\left( r\right) }N\left( r\right) }{{r}^{2}{\left( {q}^{3} + {r}^{3}\right) }^{2}}\left\lbrack  {{\left( {q}^{3} + {r}^{3}\right) }^{2} \left(1-N\left(r\right)\right)}\right. \notag \\
 & \quad 
 +12{q}^{3}M{r}^{3}{\xi}^{\prime }\left( \phi \right){\phi }^{\prime }\left( r\right)-\left(2{q}^{6}{r}^{2}+4{q}^{3}{r}^{5}+2{r}^{8}\right){\phi }^{\prime }{\left( r\right) }^{2} \notag \\
 & \quad \left. {+6{q}^{3}M{r}^{2}\xi\left( \phi \right)\left(-1+2{r}^{2}{\phi }^{\prime }{\left( r\right) }^{2}\right)+3{q}^{3}M{r}^{2}{\xi}^{\prime \prime }\left( \phi \right)}\right\rbrack \text{ . }
\end{align}

\begin{figure*}[t!]
   \centering
\subfigure[$\xi\left( \phi \right)  = 1 - \alpha {\phi }^{2}$]{\includegraphics[width=0.4\textwidth]{ 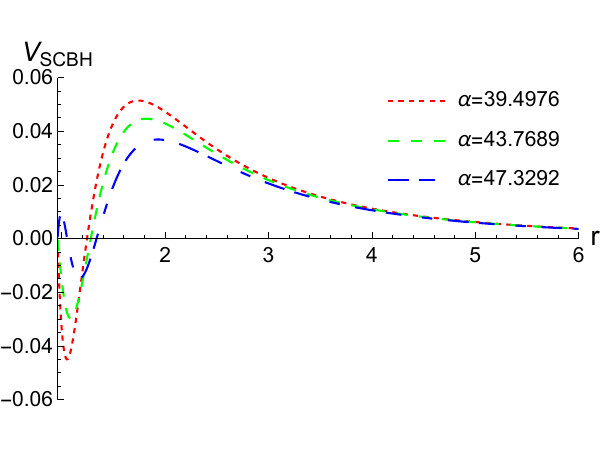}}
\qquad
\subfigure[$\xi\left( \phi \right)  = {e}^{-\alpha {\phi }^{2}}$]{\includegraphics[width=0.4\textwidth]{ 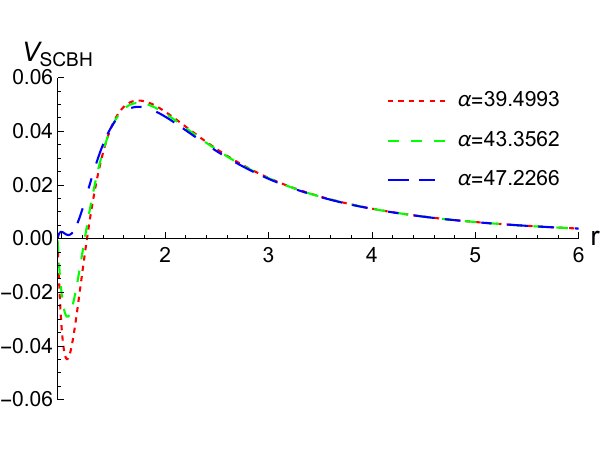}}
\caption{Three scalar potentials ${V}_{\text{ SCBH }}$ for the $l = 0$ scalar mode on the $n = 0$ branch. }\label{fig4}
\end{figure*}

\begin{figure*}[t!]
   \centering
\subfigure[$\xi\left( \phi \right)=1 - \alpha {\phi }^{2}$]{\includegraphics[width=0.4\textwidth]{ 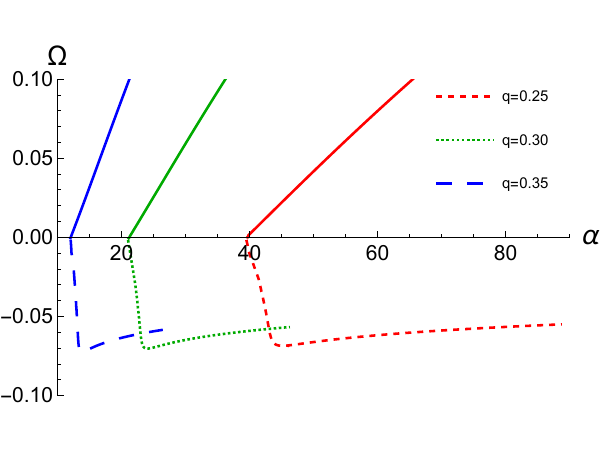}}
\qquad
\subfigure[$\xi\left( \phi \right)={e}^{-\alpha {\phi }^{2}}$]{\includegraphics[width=0.4\textwidth]{ 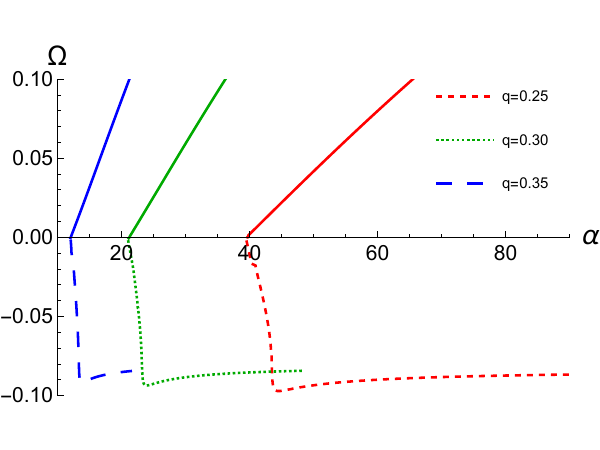}}
\caption{The eigenvalue $\Omega$ is shown as a function of $\alpha$ for the $l = 0$ scalar mode along the $n = 0$ branch, indicating stability. Three dotted curves start from ${\alpha }_{n = 0} = {39.4976}$, ${21.1872}$, and ${12.0436}$. Three solid lines denote the unstable Hayward black holes [see Fig.\ref{fig2}]. }\label{fig5}
\end{figure*}

For the quadratic coupling, the effective potentials along the $n=0$ branch are shown in Fig.~\ref{fig4}(a). Although a shallow negative region appears near the horizon for some parameter values, this feature alone is insufficient to establish an instability. The relevant criterion is the sign of the eigenvalue $\Omega$: modes with $\Omega>0$ grow exponentially in time, whereas modes with $\Omega<0$ correspond to decaying perturbations. We therefore solve Eq.~\eqref{radialeq} numerically, imposing the appropriate boundary conditions at the horizon and at spatial infinity. The results displayed in Fig.~\ref{fig5}(a) show that the fundamental branch remains stable against radial ($l=0$) scalar perturbations for the values of $q$ considered here.

The same analysis can be repeated for the exponential coupling. As illustrated in Fig.~\ref{fig4}(b), its effective potential has a profile very similar to that of the quadratic model, with the dominant support again located outside the horizon. Solving the corresponding perturbation equation with the same radial boundary conditions, we obtain the spectrum shown in Fig.~\ref{fig5}(b). The resulting eigenvalues satisfy $\Omega<0$ on the $n=0$ branch, indicating that the fundamental scalarized Hayward black holes are also radially stable in the exponential model.

\section{Conclusion and discussions}
\label{5s}

In this paper, we showed that Hayward black holes in Einstein-nonlinear electromagnetic-scalar gravity can undergo spontaneous scalarization once the scalar-free background becomes tachyonically unstable. By analyzing the corresponding scalar clouds, we identified the bifurcation points from which scalarized charged black holes emerge in both the quadratic and exponential coupling models. These solutions are organized into discrete branches classified by the node number of the scalar profile. Focusing on the fundamental branch, we found that it is radially stable under $l=0$ perturbations in the cases examined here.

Under linear test scalar perturbations, the scalar-free Hayward black hole is unstable for $\alpha>\alpha_{n=0}(q)$ and stable for $\alpha<\alpha_{n=0}(q)$. The critical value $\alpha_{n=0}(q)$ therefore plays a dual role: it marks both the onset of tachyonic instability and the bifurcation point of the fundamental scalarized branch. For the two coupling choices considered in this work, namely the quadratic and exponential models, scalarized charged black holes on the $n=0$ branch exist for $\alpha\geq\alpha_{n=0}(q)$. Moreover, the increase in $\alpha_{n=0}(q)$ as the magnetic charge decreases shows that the onset of scalarization is progressively suppressed for smaller $q$. Our radial perturbation analysis further indicates that the fundamental branch is stable, whereas the excited branches with $n\neq0$ are likely unstable. Thus, the stable $n=0$ scalarized solutions are natural candidates for the endpoint of the instability and may be relevant to future astrophysical tests \cite{Stuchlik:2019}.

\hspace*{3em}

\vspace{1cm}

{\bf Acknowledgments}

This work is supported by the National Natural Science Foundation of China (NSFC) (12305064, 12365009, 12205123), and by the Jiangxi Provincial Natural Science Foundation (20224BAB211020, 20232BAB201039, 20232BAB211029).


\vspace{1cm}



\begin{thebibliography}{}

\bibitem{Hawking:1973}
S. W. Hawking and G. F. R. Ellis, Cambridge University Press, Cambridge (1973).
\bibitem{Bardeen:1968}
J. Bardeen,  In Proc. Int. Conf. GR5, Tbilisi,
 volume 174, 1968.
\bibitem{Hayward:2005gi}
S.~A.~Hayward,
Phys. Rev. Lett. \textbf{96} (2006), 031103
[arXiv:gr-qc/0506126 [gr-qc]].
\bibitem{Fan:2016rih}
Z.~Y.~Fan,
Eur. Phys. J. C \textbf{77}, no.4, 266 (2017)
[arXiv:1609.04489 [hep-th]].
\bibitem{Fan:2016hvf}
Z.~Y.~Fan and X.~Wang,
Phys. Rev. D \textbf{94}, no.12, 124027 (2016)
[arXiv:1610.02636 [gr-qc]].
\bibitem{Huang:2025uhv}
H.~Huang and X.~P.~Rao,
Phys. Rev. D \textbf{111}, no.10, 104040 (2025)
[arXiv:2503.13133 [gr-qc]].

\bibitem{Ayon-Beato:1998}
E. Ay\'on-Beato and A. Garcia, Phys. Rev. Lett. {\bf80}, 5056 (1998).
\bibitem{Ayon-Beato:2000}
E. Ay\'on-Beato and A. Garcia, Phys. Lett. B {\bf493}, 149 (2000).

\bibitem{Huang:2020}
Y.~Huang, Q.~Pan, W.~L.~Qian, J.~Jing and S.~Wang,
Sci. China Phys. Mech. Astron. \textbf{63}, no.3, 230411 (2020)
doi:10.1007/s11433-019-9604-x

\bibitem{Zou:2021}
Y.~Zou, M.~Wang and J.~Jing,
Sci. China Phys. Mech. Astron. \textbf{64}, no.5, 250411 (2021)
doi:10.1007/s11433-020-1674-0

\bibitem{Berej:2006cc}
W.~Berej, J.~Matyjasek, D.~Tryniecki and M.~Woronowicz,
Gen. Rel. Grav. \textbf{38}, 885-906 (2006)
[arXiv:hep-th/0606185 [hep-th]].
\bibitem{Kumara:2021hlt}
A.~N.~Kumara, S.~Punacha, K.~Hegde, C.~L.~A.~Rizwan, K.~M.~Ajith and M.~S.~Ali,
Int. J. Mod. Phys. A \textbf{38}, no.29n30, 2350151 (2023)
[arXiv:2106.11095 [gr-qc]].


\bibitem{Damour:1993}
T. Damour and G. Esposito-Far\`ese, Phys. Rev. Lett. {\bf70}, 2220 (1993).
\bibitem{Damour:1996}
T. Damour and G. Esposito-Far\`ese, Phys. Rev. D {\bf54}, 1474 (1996).


\bibitem{Doneva:2017bvd}
D.~D.~Doneva and S.~S.~Yazadjiev,
Phys. Rev. Lett. \textbf{120}, no.13, 131103 (2018)
[arXiv:1711.01187 [gr-qc]].
\bibitem{Antoniou:2017acq}
G.~Antoniou, A.~Bakopoulos and P.~Kanti,
Phys. Rev. Lett. \textbf{120}, no.13, 131102 (2018)
[arXiv:1711.03390 [hep-th]].
\bibitem{Silva:2017uqg}
H.~O.~Silva, J.~Sakstein, L.~Gualtieri, T.~P.~Sotiriou and E.~Berti,
Phys. Rev. Lett. \textbf{120}, no.13, 131104 (2018)
[arXiv:1711.02080 [gr-qc]].


\bibitem{Herdeiro:2018wub}
C.~A.~R.~Herdeiro, E.~Radu, N.~Sanchis-Gual and J.~A.~Font,
Phys. Rev. Lett. \textbf{121}, no.10, 101102 (2018)
[arXiv:1806.05190 [gr-qc]].
\bibitem{Fernandes:2019rez}
P.~G.~S.~Fernandes, C.~A.~R.~Herdeiro, A.~M.~Pombo, E.~Radu and N.~Sanchis-Gual,
Class. Quant. Grav. \textbf{36}, no.13, 134002 (2019)
[erratum: Class. Quant. Grav. \textbf{37}, no.4, 049501 (2020)]
[arXiv:1902.05079 [gr-qc]].
\bibitem{Myung:2018vug}
Y.~S.~Myung and D.~C.~Zou,
Eur. Phys. J. C \textbf{79}, no.3, 273 (2019)
[arXiv:1808.02609 [gr-qc]].
\bibitem{Zou:2019bpt}
D.~C.~Zou and Y.~S.~Myung,
Phys. Rev. D \textbf{100}, no.12, 124055 (2019)
[arXiv:1909.11859 [gr-qc]].
\bibitem{Astefanesei:2019pfq}
D.~Astefanesei, C.~Herdeiro, A.~Pombo and E.~Radu,
JHEP \textbf{10}, 078 (2019)
[arXiv:1905.08304 [hep-th]].

\bibitem{Cunha:2019}
P. V. Cunha, C. A. Herdeiro, and E. Radu, Phys. Rev. Lett. {\bf123}, 011101 (2019).
\bibitem{Collodel:2020}
L. G. Collodel, B. Kleihaus, J. Kunz, and E. Berti, Class. Quantum Grav. {\bf37}, 075018 (2020).

\bibitem{Dima:2020}
A. Dima, E. Barausse, N. Franchini, and T. P. Sotiriou, Phys. Rev. Lett. {\bf125}, 231101 (2020).
\bibitem{DonevaPRD:2020}
D. D. Doneva, L. G. Collodel, C. J. Krüger, and S. S. Yazadjiev, Phys. Rev. D {\bf102}, 104027 (2020).
\bibitem{DonevaEPJC:2020}
D. D. Doneva, L. G. Collodel, C. J. Krüger, and S. S. Yazadjiev, Eur. Phys. J. C {\bf80}, 1205 (2020).
\bibitem{Herdeiro:2021}
C. A. R. Herdeiro, E. Radu, H. O. Silva, T. P. Sotiriou, and N. Yunes, Phys. Rev. Lett. {\bf126}, 011103 (2021).
\bibitem{Berti:2021}
E. Berti, L. G. Collodel, B. Kleihaus, and J. Kunz, Phys. Rev. Lett. {\bf126}, 011104 (2021).

\bibitem{Whinnett:2000}
A. W. Whinnett, Phys. Rev. D {\bf61}, 124014 (2000).
\bibitem{Brihaye:2019}
Y. Brihaye and B. Hartmann, J. High Energy Phys. {\bf 09}, 049 (2019).
\bibitem{Minamitsuji:2021}
M. Minamitsuji and S. Tsujikawa, Phys. Lett. B {\bf 820}, 136509 (2021).


\bibitem{PengJHEP} Y. Peng, J. High Energy Phys. {\bf 12}, 064 (2019).

\bibitem{Zhang:2024bfu}
L.~Zhang, Q.~Pan, Y.~S.~Myung and D.~C.~Zou,
Phys. Rev. D \textbf{110}, no.12, 124036 (2024)
[arXiv:2409.11669 [gr-qc]].


\bibitem{Benavides-Gallego:2018odl}
C.~A.~Benavides-Gallego, A.~A.~Abdujabbarov and C.~Bambi,
Phys. Rev. D \textbf{101}, no.4, 044038 (2020)
[arXiv:1811.01562 [gr-qc]].
\bibitem{Nam:2018uvc}
C.~H.~Nam,
Gen. Rel. Grav. \textbf{50}, no.6, 57 (2018)
\bibitem{Li:2025glq}
Q.~Q.~Li, Y.~Zhang and H.~Iminniyaz,
Chin. Phys. C \textbf{49}, no.10, 105106 (2025)
[arXiv:2501.15983 [gr-qc]].

\bibitem{LiLi:2025qgo}
L.~Li, Z.~Sun and F.~G.~Yang,
doi:10.1103/l4tf-kqyl
[arXiv:2512.19377 [gr-qc]].

\bibitem{Stuchlik:2019}
Z. Stuchl\'\i{}k and J. Schee, Eur. Phys. J. C {\bf79},  44 (2019).



\end{thebibliography}
\end{document}